\documentclass[journal=nalefd,manuscript=article]{achemso}

\usepackage[version=3]{mhchem} 
\usepackage[utf8]{inputenc}
\usepackage{setspace}
\usepackage{amsmath}
\usepackage{amssymb}
\usepackage{csquotes}
\usepackage{graphicx}
\usepackage{subfig}
\usepackage{minibox}
\usepackage{wrapfig}
\usepackage{array}
\usepackage{url}
\usepackage[T1]{fontenc}
\author{Roop K.Mech}
\affiliation {\small \textit Institute of Condensed Matter and Nanosciences, Université Catholique de Louvain (UCLouvain), 1348 Louvain-la-Neuve, Belgium}

\author{Jean Spièce}
\affiliation {\small \textit Institute of Condensed Matter and Nanosciences, Université Catholique de Louvain (UCLouvain), 1348 Louvain-la-Neuve, Belgium}

\author{Kenji Watanabe}
\affiliation {\small \textit National Institute for Materials Science,  1-1 Namiki, Tsukuba 305-0044, Japan}

\author{Takashi Taniguchi}
\affiliation {\small \textit National Institute for Materials Science,  1-1 Namiki, Tsukuba 305-0044, Japan}

\author{Pascal Gehring}
\email{pascal.gehring@uclouvain.be}
\affiliation {\small \textit Institute of Condensed Matter and Nanosciences, Université Catholique de Louvain (UCLouvain), 1348 Louvain-la-Neuve, Belgium}

\title{Versatile polymer method to dry-flip two-dimensional moiré hetero structures for nanoscale surface characterization}

\abbreviations{}

\begin{document}
\captionsetup[figure]{labelfont={bf},labelformat={default},labelsep=period,name={Figure}}

\begin{abstract} 
The recent discovery of magic angle twisted bilayer graphene (MATBG), in which two sheets of monolayer graphene are precisely stacked to a specific angle, has opened up a plethora of new opportunities in the field of topology, superconductivity, and other strongly correlated effects. Most conventional ways of preparing twisted bilayer devices require the use of high process temperatures and solvents and are not well-suited for preparing samples which need to be flipped to be compatible with characterization techniques like STM, ARPES, PFM, SThM etc. Here, we demonstrate a very simple polymer-based method using Polyvinyl Chloride (PVC), which can be used for making flipped twisted bilayer graphene devices. This allowed us to produce flipped twisted samples without the need of any solvents and with high quality as confirmed by Piezoresponse Force Microscopy. We believe that this dry flip technique can be readily extended to twist 2D materials beyond graphene, especially air-sensitive materials which require operation under inert atmosphere, where often solvents cannot be used.
\end{abstract}

\section{Introduction}
The controlled assembly of single sheets of two-dimensional (2D) materials into heterostructures has paved the way for fabricating novel designer materials for various applications including electronics\cite{lin2015atomically}, memory storage\cite{lei2015optoelectronic}, sensing\cite{loan2014graphene}, and photovoltaics\cite{furchi2014photovoltaic}. Stacking 2D materials with a controlled angle between the mono-atomic layers has also opened new doors for exploring novel physics phenomena such as topology\cite{zhang2022two}, superconductivity\cite{fatemi2018electrically}, magnetism\cite{klein2018probing}, and strongly correlated physics\cite{cao2018correlated}. The most established method for building 2D heterostructures involves using adhesive polymer stamps made of e.g. poly-carbonate (PC)\cite{purdie2018cleaning} and polypropylene carbonate (PPC)\cite{kinoshita2019dry} to pick up 2D materials from arbitrary substrates. PPC is useful for picking up 2D materials between 40°C - 70°C and PC is used at high temperatures between 100°C - 110°C. The final stack is then released onto a substrate by raising the temperature above the glass transition temperature of the polymer (e.g. 180°C for PC) followed by a cleaning step in organic solvents (typically Chloroform for PC). This technique has proven effective for making 2D heterostructures for transport measurements, where the material under test is typically fully encapsulated in between hexagonal boron nitride layers. However, it has limited suitability for fabricating samples for e.g. scanning probe techniques such as STM, SThM, PFM, ARPES, and MFM where a non-encapsulated 2D material with exposed surface is required. Previous studies\cite{cao2020movable,oh2021evidence} have achieved this by flipping over the van der Waals (vdW) heterostructure using PVA. However, flipping the heterostructure using PVA requires water as a solvent, which is not suitable for air- and solvent-sensitive 2D materials. In this study, we have developed a simple, reliable and fully dry method for picking up, flipping over, and releasing 2D heterostructures using Polyvinyl Chloride (PVC). PVC allows for easy pick-up and release of 2D materials by adjusting the temperature without the need for solvents. Additionally, the heterostructure can be flipped by transferring from a thicker PVC polymer to a thinner PVC polymer, making PVC highly effective in designing 2D architectures for both transport and scanning probe techniques. The various polymer methods used in van der Waals stacking of 2D materials are listed in Table \ref{tab:tab1}. 

\begin{table*}[h]
  \caption{containing various polymer method mentioning pick up temperature of each method, polymer cleaning method and their efficiency in flipping 2D stack. Polypropylene carbonate (PPC)\cite{wang2013one,yankowitz2019tuning,xu2019ws2}, Polycarbonate (PC)\cite{cao2018unconventional,cao2018correlated,agarwal2023ultra,kim2022evidence}, Polydimethylsiloxane (PDMS)\cite{merino2023two,khazaeinezhad2014direct,li2022constructing,jain2018minimizing}, Polyvinyl alcohol (PVA)\cite{van2016pmma,li2022poly}, Polyvinyl Chloride (PVC)$^*$ and Polycaprolactone (PCL)\cite{son2020strongly,choi2021high}. $^*$This work
  }
    \centering
    \resizebox{1.0\textwidth}{!}
    {
    \begin{tabular}{|c|m{4cm}|c|m{4cm}|c|m{3cm}|c|m{4cm}|}
    \hline      
   \textbf{Polymer Method} & \textbf{Pick up Temperature} & \textbf{Method to remove polymer} &\textbf{Efficient in flipping 2D heterostructure} \\ 
    \hline 
    PPC & 40ºC - 70ºC &Dichloromethane or Acetone & No \\
     \hline 
     PC & 90ºC - 110ºC & Chloroform & No \\
     \hline
     PDMS &	pick up impossible & Dry release &	No \\
     \hline
     PVA	& 90ºC	& Water &	Yes, but needs water as solvent \\
     \hline
     PVC$^*$ & 25ºC - 35ºC (thick) 70 ºC - 90 ºC (thin) &	Dry release	& Yes \\
     \hline
     PCL	& 50ºC - 60ºC & Tetrahydrofuran(THF) or vacuum annealing &	No \\
         \hline
    \end{tabular}
    }
    \label{tab:tab1}
    
 \end{table*}

\begin{figure}[!h]
\centering \includegraphics[width=1\textwidth]{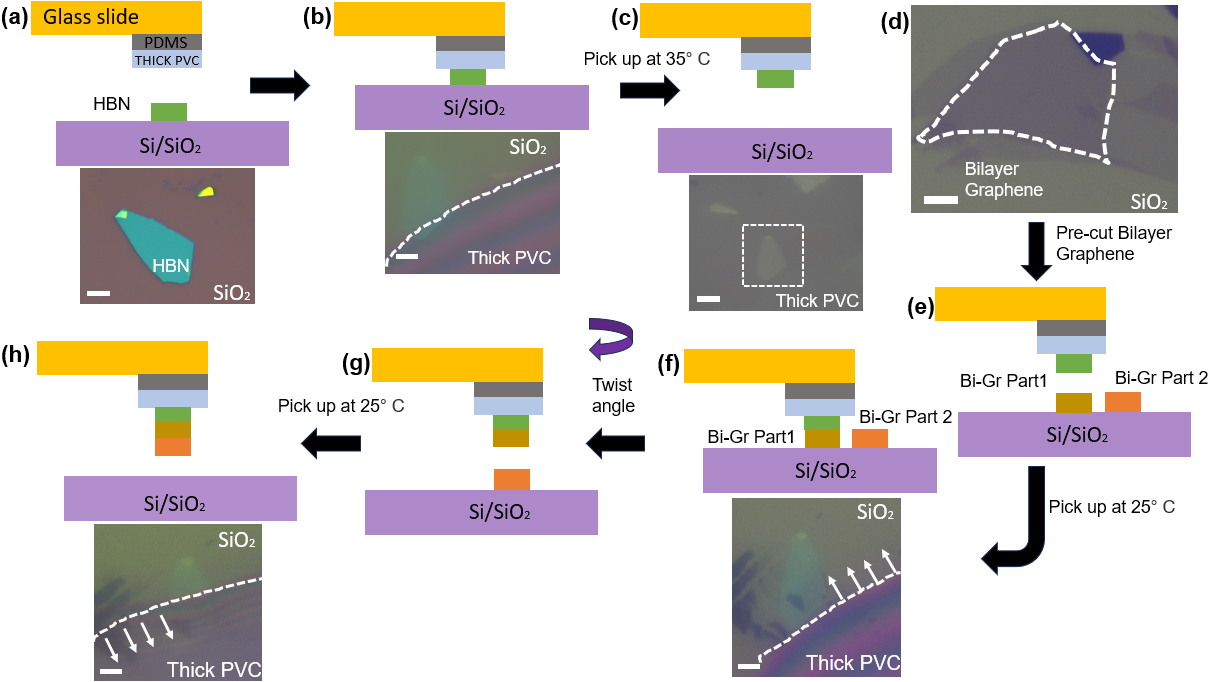}
\caption{\textbf{Schematic and optical images of the method developed for stacking of twisted double bilayer device using thick PVC/PDMS stamp.} (a-c) A thick PVC/PDMS stamp is used to pick up hBN flake at 35°C. (d) Bi-layer graphene is exfoliated onto SiO$_2$/Si and is pre-cut using an AFM tip. (e-f) The first part of the pre-cut bi-layer graphene is picked up at 25°C. (g-h) Twist angle is applied and the rotated second part of the pre-cut bi-layer graphene is picked up at 25°C. All related optical images are displayed below their respective steps. Scale bars:10 µm.
}
\label{fig1}
\end{figure}

\section{Results and discussions}
 Figure \ref{fig1} illustrates the schematic of the transfer method that we use to fabricate flipped twisted bilayer devices. Hexagonal boron nitride (HBN) and graphene were exfoliated on SiO$_2$/Si using the standard Scotch tape method. The graphene was pre-cut at 100°C using an AFM tip. A thick PVC stamp was used to pick up hBN at a temperature of 35°C (see Figure \ref{fig1} a-c). One half of the pre-cut graphene flake was then picked up by the hBN/PVC/PDMS stamp at room temperature (25°C, Figure 1d-f). Subsequently, the second half of the pre-cut graphene was rotated at a small angle and then picked up by the Gr1/hBN/PVC/PDMS stamp at 25°C (Figure \ref{fig1}g-h). The low pick-up temperature of PVC, compared to PC, prevents twist angle relaxation, providing an additional advantage for twisted materials. Such heterostructure could be readily used for device fabrication. Furthermore, it is worth to mention that the PVC polymer can also pick up air sensitive 2D materials directly from the substrate without the need of hBN and can release them "dry" on substrate without the need of any solvent. To demonstrate this, we used PVC polymer to pick up and dry-release air- sensitive 2D flakes of In$_2$Se$_3$ , CrTe$_2$ and NiI$_2$ inside a Nitrogen filled glovebox (see Supplementary Figures S1 and S2). 
\begin{figure}[!h]
\centering \includegraphics[width=1\textwidth]{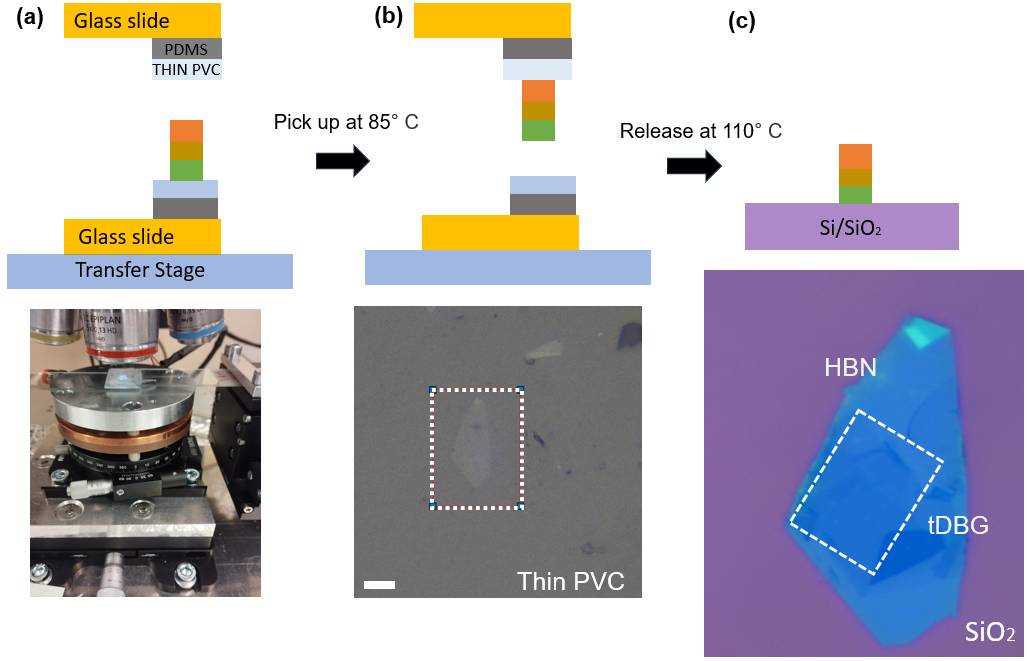}
\caption{\textbf{Schematic and optical images of the flipping operation of twisted double bilayer graphene using PVC/PDMS stamp.} (a) The thick PVC containing the twisted double bilayer graphene placed on the transfer stage and it is approached by a thin PVC/PDMS stamp.(b) The thin PVC/PDMS is used to pick up twisted double bilayer graphene at 85°C. (c) The flipped stack is finally released on O$_2$ plasma etched SiO$_2$/Si without the use of any solvent. All related optical images are displayed below their respective steps. Scale bars: 10 µm.
}
\label{fig2}
\end{figure}

Next, we show how to flip the twisted bilayer heterostructure. This is achieved by using a thinner PVC polymer: for PVC, the adhesion of thick films decreases faster with increasing temperature than that of thin films. As an example, for the thick PVC used in this study, the adhesion between the thicker PVC and the 2D material is significantly reduced at temperatures around 50-70°C, while the thin PVC used here has strong adhesion (pick-up temperature) between  80-90°C. This allows to transfer the material from one polymer to the other. Figure \ref{fig2} illustrates the schematic of the flip and release operation for twisted devices. The 2D heterostructure on the thicker PVC is placed on the transfer stage and gently approached by the thinner PVC at 70°C (Figure \ref{fig2}a). Then, the temperature is increased to 85°C to pick up the heterostructure (Figure \ref{fig2}b). Finally, the flipped twisted material is released onto a SiO$_2$/Si substrate at a temperature of 110°C (Figure \ref{fig2}c).

\begin{figure}[!h]
\centering \includegraphics[width=1\textwidth]{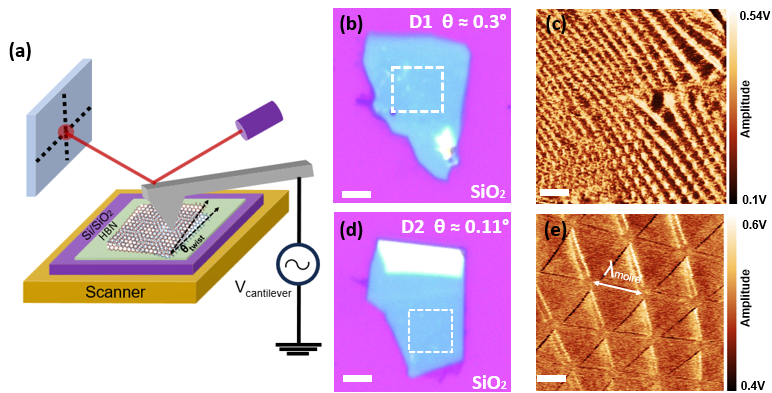}
\caption{\textbf{Characterization of moiré pattern in twisted bilayer graphene (TBG).} (a) Schematic of the experimental setup used for Piezoresponse Force Microscopy (PFM). The optical images (b,d) and PFM amplitude (c,e) of minimally twisted TBG devices with twist angles of 0.3$^\circ$ (D1) and 0.1$^\circ$ (D2),  respectively.          
Scale bars: 10 µm (b) and (d) , 1 µm (c) and 400 nm (e).}
\label{fig3}
\end{figure}

In the following, we demonstrate that the flipped heterostructures are suitable for scanning probe microscopy by performing Piezoresponse Force Microscopy (PFM, see Figure \ref{fig3}a) experiments on them. Figure \ref{fig3}b and \ref{fig3}d display the optical image of two minimally twisted devices 0.3°and 0.1°, respectively. The effective twist angle is calculated using the relationship between the rotation angle $\theta$ and moir\'e wavelength $\lambda_\mathrm{moire}$: $\lambda_\mathrm{moire} = (a/2)*\mathrm{csc}(\theta/2)$, where $a$ is the lattice constant of graphene\cite{kim2017tunable}. In Figure \ref{fig3}c and \ref{fig3}e we show the PFM amplitude images of devices D1 and D2. We observe triangular pattern which we attribute to the formation of commensurate stacking domains. These domains create intralayer strain gradients which result in an electromechanical coupling to the out-of-plane electric field, allowing for their direct visualization, as previously demonstrated.\cite{mcgilly2020visualization,canetta2023quantifying,li2021unraveling}

\section{Conclusion}
In conclusion, this paper showcases a simple and efficient polymer-to-polymer transfer method for twisted 2D materials utilizing a PVC/PDMS stamp. The pick-up and release temperatures of the PVC polymer are influenced by the film thickness, enabling the flipping of twisted 2D heterostructures without the need for solvents. This PVC polymer transfer technique can be readily applied to create twisted 2D heterostructures with various other 2D materials, leading to significant progress in exploring the exotic physics of these materials.

\section{Methods}

\subsection{Piezo Force Microscopy}
PFM measurements were performed with a Digital Instruments Multimode IIIa atomic force microscope. An AC bias (between 1 and 3 V) is applied between the conductive AFM tip (ASYELEC.01-R2, Oxford Instruments) and the sample to produce an electric field along the vertical direction. This induces periodic a converse piezoelectric effect on the sample that is detected by measuring the torsion or deflection of the AFM lever via the AFM laser at the bias frequency via lock-in detection (MFLI, Zurich Instruments). In this work, we used the first contact resonance to enhance the signal to noise ration. The typical contact resonance lies between 270 kHz and 320 kHz.

\begin{acknowledgement}
The authors acknowledge financial support from the F.R.S.-FNRS of Belgium (FNRS-CQ-1.C044.21-SMARD, FNRS-CDR-J.0068.21-SMARD, FNRS-MIS-F.4523.22-TopoBrain, FNRS-PDR-T.0128.24-ART-MULTI, FNRS-CR-1.B.463.22-MouleFrits), from the EU (ERC-StG-10104144-MOUNTAIN), from the Federation Wallonie-Bruxelles through the ARC Grant No. 21/26-116, and from the FWO and FRS-FNRS under the Excellence of Science (EOS) programme (40007563-CONNECT). K. W. and
T. T. acknowledge support from JSPS KAKENHI (Grant
Numbers 19H05790, 20H00354 and 21H05233).

\end{acknowledgement}

\bibliography{ref}
\end{document}


\maketitle

\section{Contents: }Figure S1. Fermomagnetic/Ferroelectric Junction between CrTe$_2$ and In$_2$Se$_3$. 

Figure S2. Vertical junction between Nil$_2$ and hBN.

\begin{figure}[!h]
\centering \includegraphics[width=1\textwidth]{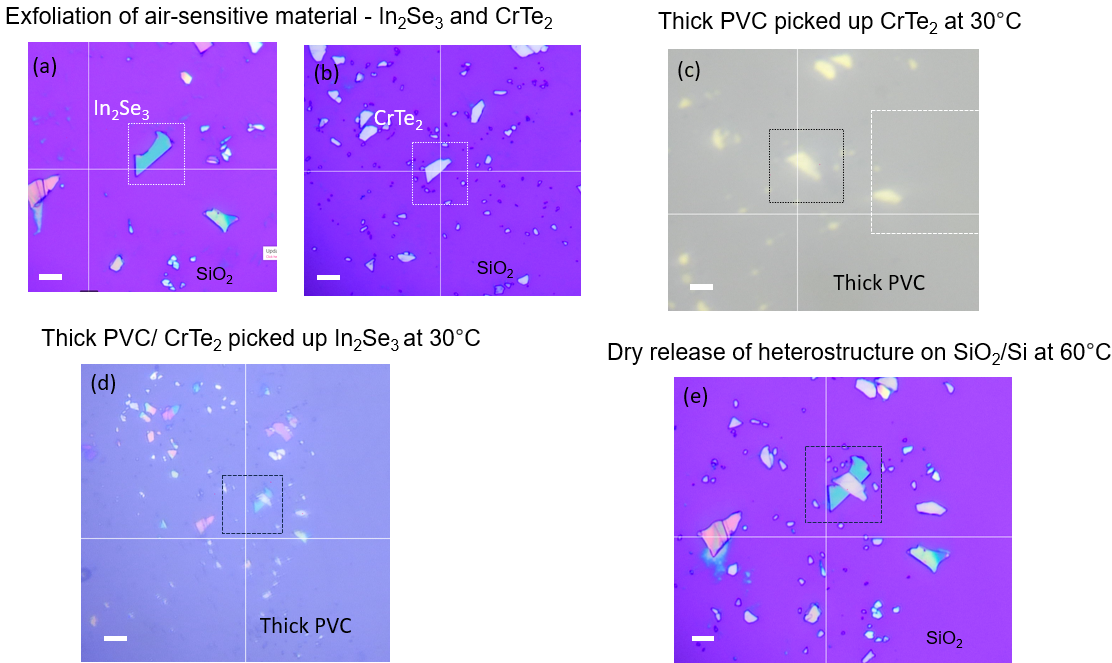}
\caption{\textbf{Optical image of a Fermomagnetic (FM)/Ferroelectric(FE) Junction made between CrTe$_2$(FM) and In$_2$Se$_3$(FE). Scotch tape exfoliation of (a) In$_2$Se$_3$ and (b) CrTe$_2$ on SiO$_2$/Si. (c) Picking up of CrTe$_2$ with a thick PVC/PDMS stamp at 30°C. (d) In$_2$Se$_3$ is picked up by CrTe$_2$/thick PVC/PDMS at  30°C. (e) Finally the heterostructure is dry released on SiO$_2$/Si at 60°C.}. Scale bars: 10 µm (a),(b),(c) and (e) , 20 µm (d). 
}
\label{S1}
\end{figure}
 
\begin{figure}[!h]
\centering \includegraphics[width=1\textwidth]{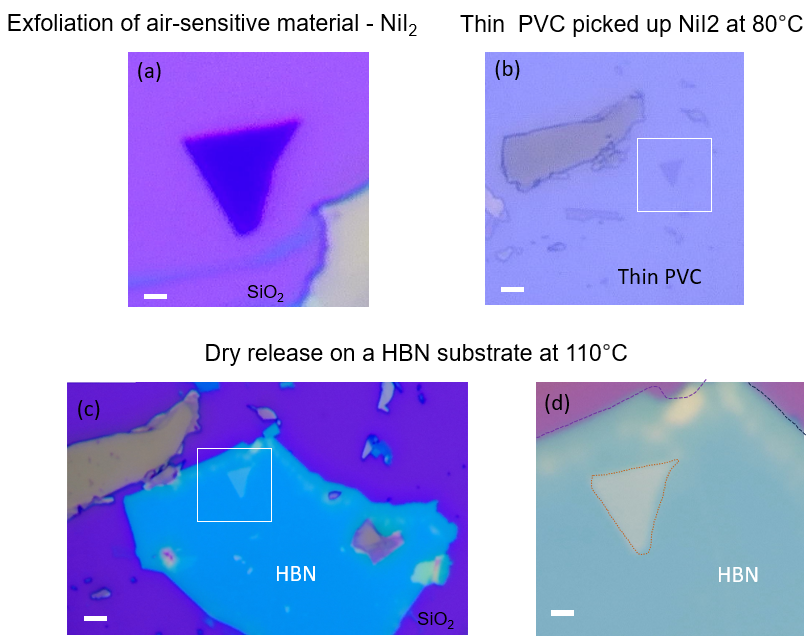}
\caption{\textbf{Optical image of a vertical junction between Nil$_2$ and hBN.(a) Scotch tape exfoliation of thin flake of Nil$_2$ on SiO$_2$/Si. (b) Picking up Nil$_2$ with a thin PVC/PDMS stamp at 80°C. (c) Dry releasing of Nil$_2$ on HBN at 110°C. (d) Zoom image of Nil$_2$/HBN stack.}. Scale bars: 5 µm (a) and (d), 20 µm (b) and (c). 
}
\label{Figure S1}

\end{figure}